\documentclass[%
twocolumn,
superscriptaddress,
nofootinbib,
 amsmath,amssymb,
 aps, prd
]{revtex4-2}

\usepackage{graphicx}
\usepackage{tensor}
\usepackage{siunitx}
\usepackage{xcolor} 
\usepackage[normalem]{ulem} 
\usepackage{lipsum} 

\usepackage[pdfborder={0 0 0}, plainpages=false, hyperfigures=true]{hyperref}

\newcommand{\beq}{\begin{equation}}
\newcommand{\eeq}{\end{equation}}
\newcommand{\beqn}{\begin{eqnarray}}
\newcommand{\eeqn}{\end{eqnarray}}

\newcommand{\lo}{\mathrel{\raise.3ex\hbox{$<$}\mkern-14mu
    \lower0.6ex\hbox{$\sim$}}}
\newcommand{\go}{\mathrel{\raise.3ex\hbox{$>$}\mkern-14mu
    \lower0.6ex\hbox{$\sim$}}}

\newcommand{\mN}{\mathcal N}

\newcommand{\UNH}{\affiliation{Department of Physics \& Astronomy, University of New Hampshire, 9 Library Way, Durham NH 03824, USA}}

\begin{document}
\title{On the difficulty of capturing the distribution function of neutrinos in neutron star merger simulations}
\author{Francois Foucart}\UNH

\begin{abstract}
The collision of two neutron stars is a rich source of information about nuclear physics. In particular, the kilonova signal following a merger can help us elucidate the role of neutron stars in nucleosynthesis, and informs us about the properties of matter above nuclear saturation. Approximate modeling of neutrinos remains an important limitation to our ability to make predictions for these observables. Part of the problem is the fermionic nature of neutrinos. By the exclusion principle, the expected value $f_\nu$ for the number of neutrinos in a quantum state is at most 1. Any process producing neutrinos is suppressed by a blocking factor $(1-f_\nu)$. Recent simulations focused on neutrino physics mostly use a gray two-moment scheme to evolve neutrinos. This evolves integrals of $f_\nu$ over momentum space, preventing direct calculations of blocking factors. Monte Carlo methods may be an attractive alternative, providing access to the full distribution of neutrinos. Their current implementation is however inadequate to estimate $f_\nu$: in our most recent simulations, a single Monte Carlo packet causes, in the worst cases, estimates of $f_\nu$ to jump from $f_\nu=0$ to $f_\nu\sim 10^5$. While this is concerning, this brazen violation of the fermionic nature of neutrinos has been largely inconsequential as the interactions used in simulations avoid direct calculations of $f_\nu$. We are however reaching a level of modeling at which this problem can no longer be ignored. Here, we discuss the relatively simple origin of this issue. We then show that very rough estimates of $f_\nu$ can in theory be obtained in merger simulations, but that they will require a combination of unintuitive weighting schemes for Monte Carlo packets and smoothing of the neutrino distribution at coarser resolution than what the merger simulation uses.
\end{abstract}

\maketitle

\section{Introduction}

Neutrino-matter interactions play a crucial role in the modeling of the merger and post-merger evolution of neutron star binaries. While not directly important to the dynamics of the merger on short time scales, neutrinos are crucial to understand the thermodynamics of the post-merger evolution and the composition of the matter ejected during and after the collision~\cite{2010CQGra..27k4107S,Deaton2013,Wanajo2014}. The latter is particularly important as it greatly impacts the properties of the UV-optical-infrared transient following the merger ({\it kilonova}), as well as the outcome of nucleosynthesis in the ejecta~\cite{Wanajo2014,2013ApJ...775...18B}.

Accurate evolution of neutrinos runs into two major issues. One is that neutrinos are trapped inside hot neutron stars, nearly free streaming in low-density matter outflows, and partially coupled in dense accretion disks. Accordingly, they require the evolution in time of Boltzmann's equation of radiation transport for the distribution function of neutrinos $f_{\nu}$, the expected occupation number of quantum states in a given region of phase space. The distribution function depends on the position, energy, and direction of propagation of the neutrinos. It is a 6-dimensional function for each species of neutrinos. This high dimensionality makes direct discretization of the problem onto a computational grid very costly without further approximations. The second issue is the complexity of neutrino-matter and neutrino-neutrino interactions. These interactions can act extremely fast in hot regions, creating stiff source terms for Boltzmann's equation and potential numerical instabilities. They also couple non-linearly the distribution function of neutrinos with different momenta and neutrinos of different species (e.g. through neutrino-antineutrino annihilation). Attempting to capture flavor conversion can increase the complexity of the problem even further, requiring us to go beyond Boltzmann transport and into quantum kinetics (see e.g.~\cite{Foucart:2022bth} for a review of neutrinos in neutron star mergers).

Recent neutron star merger simulations focused on capturing neutrino-matter interactions have relied on one of two distinct transport methods: either a gray (energy-integrated) two-moment formalism~\cite{shibata:11,Wanajo2014,Foucart:2016rxm,Radice:2021jtw}, in which we evolve moments of the distribution function (energy density, momentum flux) integrated over neutrino energies, or Monte Carlo methods~\cite{Foucart:2021mcb,Miller:2019dpt,Kawaguchi:2024naa}, in which the distribution function of neutrinos is statistically sampled by `packets' each representing a large number of neutrinos. Moments methods require the use of an approximate analytical closure for higher moments of the neutrino distribution function, which is likely inaccurate in low-density regions. They also suffer from the lack of energy and momentum space information about the distribution function when calculating neutrino-matter interactions. Monte Carlo methods, in theory, capture the full 6-dimensional distribution function of neutrinos. However, they require an approximate treatment of neutrino-matter interactions in regions where neutrinos are strongly coupled to the matter (where moments methods excel), and suffer from significant shot noise and a slow convergence rate to the correct solution. For the limited neutrino physics currently implemented in merger simulations, the two methods appear to agree within $\sim (10-20)\%$ relative errors on the main observables: luminosity, outflow masses and composition, remnant properties~\cite{Foucart:2021mcb,Foucart:2024npn}. This is likely a smaller source of error than missing physics in the simulations, e.g. flavor conversion~\cite{Qiu:2025kgy}, interactions with muons~\cite{Ng:2024zve}, neutrino-antineutrino pair production and annihilation~\cite{Fujibayashi:2017puw}, scattering of neutrinos on electrons~\cite{Cheong:2024buu}. How to implement more advanced neutrino physics for each of these two methods is an active area of research. We also note that methods combining Monte Carlo and moments algorithms show some promises~\cite{Foucart:2017mbt,Izquierdo:2023fub}, but have yet to be applied to full merger simulations.

Direct, on-the-fly computation of the distribution function of neutrinos would be extremely valuable for the implementation of more advanced neutrino physics. First, neutrinos are fermions and thus subject to Pauli's exclusion principle. Their distribution function has to satisfy $0<f_\nu<1$ (no more than one neutrino per state). When calculating neutrino matter interactions, the production of neutrinos is then suppressed by a blocking factor $(1-f_\nu)$. For some reactions (charged current reactions, elastic scattering), it is possible to rewrite the interaction terms to avoid on-the-fly calculations of $f_\nu$ (see e.g.~\cite{Burrows2006b}). For others (inelastic scattering, pair processes), no such method is available. This is the main reason pair productions are only very approximately accounted for in merger simulations, while inelastic scattering on electrons is generally ignored. Neutrino flavor conversions are also very sensitive to the angular distribution of neutrinos; the fast flavor instability for example, which may have an important impact on post-merger remnants, occurs when the net angular lepton flux changes signs as we vary the angular direction of the neutrinos~\cite{PhysRevD.84.053013,Wu:2017drk}.

In a gray two-moment scheme, the angular and spectral information is explicitly integrated over, and thus inevitably lost. Models approximating that information from the evolved moments are needed to estimate $f_\nu$. Monte Carlo methods, in theory, offer more hope. We will show here that current merger simulations remain nonetheless very far from being able to resolve $f_\nu$ (at least from single-cell information), or even from satisfying the basic requirements of the exclusion principle. With a more careful choice of the weighting of neutrino packets, we may come closer to at least a rough approximation of $f_\nu$ (errors of $\sim 0.1-0.3$ in $f_\nu$ in regions where $f_\nu \sim 1$...) as a function of energy without dramatic increases to the cost of simulations. However, this requires an unintuitive weighting of neutrino packets that may lead to increased shot noise in the coupling to the fluid. Obtaining similarly coarse angular information, integrated over neutrino energies, is an only slightly simpler task, while combining both would inevitably cause a drastic increase of the cost of Monte Carlo simulations.

Before getting into more detailed calculations, it is worth noting that the difficulty of capturing $f_\nu$ in Monte Carlo simulations should not be in any way surprising. Consider that current 3D Monte Carlo simulations of neutron star mergers use tens to hundreds of packets per cell in the best resolved regions. At the same time, low energy neutrinos have equilibrium distribution functions $f_{\rm eq} \sim O(1)$ in at least some low energy bins. Finally, if the expected number of packets in a region of phase space is $N$, the Monte Carlo sampling noise is $\sim\sqrt{N}$. Getting an accuracy of $\sim 0.1$ on $f_\nu$ when $f_\nu\sim 1$ thus requires $\sim 100$ packets within the region of phase space used to estimate $f_\nu$. Obtaining either spectral or very coarse angular information to that accuracy thus requires at least $O(10^3)$ of packets, if the packets are ideally distributed for that purpose... and we will see that in current simulations, they very much are not. This is comparable to the effective number of packets per cell available to compute $f_\nu$ in the axisymmetric simulations of~\cite{Kawaguchi:2024naa}, indicating that such accuracy is attainable in similar simulations, in theory at least. In full 3D simulations of neutron star remnants~\cite{Foucart:2021mcb} and post-merger accretion disks~\cite{Miller:2019dpt}, getting to errors below $0.3$ in $f_\nu$ would likely require significant increases in the number of packets used, as well as changes to the weighing of the packets, even if we use as little as $O(10)$ bins to discretize $f_\nu$ in momentum space. Alternatively, one could get more accurate values of an estimated $f_\nu$ by effectively averaging $f_\nu$ over a larger region of phase space (a larger spatial volume than the simulation grid cell, time averaging, and/or coarser energy discretization). A solution to this problem will likely require a combination of these methods.

\section{Background}

\subsection{Neutrino distribution function and number of available states}

Consider a 4-dimensional spacetime parametrized by a timelike variable $t$ defining spacelike $t=t_0$ slices. The distribution function $f_\nu(t,x^i,p_i)$ of a massless particle is defined so that the number of particles $N$ in a region of phase space $D$ is
\beq
N(t|D) = \frac{1}{h^3} \int_D d^3 x^i d^3 p_i f_\nu(t,x^i,p_i),
\eeq
with $x^i$ the position of the particles, $p_i$ the spatial components of the 4-momentum one-form $p_\mu$, and $h$ Planck's constant, introduced so that $f_\nu$ properly represents the expected occupation level of each quantum state. We assume that we consider regions $D$ small enough that we can reasonably define an orthonormal coordinate system in which spacetime curvature can be ignored. We also ignore the mass of neutrinos, and assume that all (anti)neutrinos have left (right)-handed chirality. The distribution function of each species of neutrinos then has to satisfy $0<f_\nu<1$. In neutron star merger simulations, the spacetime metric generally varies slowly between neighboring cells, and typical neutrino energies are $O(10\,{\rm MeV})$; these assumptions are then well justified.

In Monte Carlo simulations, the distribution function is discretized by `packets' each representing a large number of neutrinos. We write
\beq
f(t,x^i,p_i) = \sum_k n_k \delta^3(x^i - x^i_k(t)) \delta^3(p_i - p_i^k(t)),
\eeq
with the sum being over all packets, $n_k$ the number of neutrinos represented by packet $k$, $x^i_k$ the packet's position, and $p_i^k$ the spatial components of its momentum. As a sum of Dirac deltas, this function obviously does not satisfy the condition $f_\nu<1$. The Monte Carlo simulations can only be used to calculate $f_\nu$ for a choice of averaging domain $D$. Within a domain $D$ where the expectation value for the number of neutrino packets is $\mN \gg1$, the Monte Carlo simulation should have $\sim \mN \pm \sqrt{\mN}$ packets (see the Appendix for a discussion of the limits of this approximation). The larger $D$ is, the lower our sampling noise will be -- at the cost of course of the resolution at which we capture $f_\nu$.\footnote{We note that in this manuscript we deal with the number and energy of neutrinos represented by a single packet, the number and energy of all neutrinos within a region $D$ of phase space, and the number of packets used to represent those neutrinos. To differentiate these concepts more clearly, we denote the first with lower-case symbols $(n,e)$, the second with upper-case symbols $(N,E)$, and the third as $\mN$.}

A useful quantity that can be derived from the definition of $f_\nu$ is the maximum number of neutrinos $N_{\rm max}$ in $D$. If we consider neutrinos within a spacetime volume $\Delta V$, an energy range $\epsilon_0<\epsilon_\nu<\epsilon_1$ and a solid angle $\Delta \Omega$ (for the direction of $p_i$ on the unit sphere) and set $f_\nu=1$ (all states are occupied), we get
\beq
N_{\rm max}(D) = \frac{1}{(hc)^3} (\Delta \Omega) (\Delta V) \frac{\epsilon_1^3-\epsilon_0^3}{3},
\eeq
or, scaling to values typical of neutron star merger simulations,
\beq
N_{\rm max}(D) = 1.76\times 10^{46} \frac{\Delta \Omega}{4\pi} \frac{\Delta V}{(100\,{\rm m})^3} \frac{\epsilon_1^3-\epsilon_0^3}{(20\,{\rm MeV})^3}.
\eeq
The same calculation for the maximum energy $E_{\rm max}$ of neutrinos in $D$ gives
\beq
E_{\rm max}(D) = \frac{1}{(hc)^3} (\Delta \Omega) (\Delta V) \frac{\epsilon_1^4-\epsilon_0^4}{4}, 
\eeq
or
\beq
E_{\rm max}(D) = 4.22\times 10^{41}{\rm ergs} \frac{\Delta \Omega}{4\pi} \frac{\Delta V}{(100\,{\rm m})^3} \frac{\epsilon_1^4-\epsilon_0^4}{(20\,{\rm MeV})^4}.
\eeq
As energies are often expressed in units where $M_\odot=c=G=1$ in merger simulations, it will be convenient to note that this is equivalent to
\beq
E_{\rm max} =
 2.35\times 10^{-13} M_\odot c^2 \frac{\Delta \Omega}{4\pi} \frac{\Delta V}{(100\,{\rm m})^3} \frac{\epsilon_1^4-\epsilon_0^4}{(20\,{\rm MeV})^4}.
\eeq
To put into context the problems that will inevitably arise when attempting to evaluate $f_\nu$ within a cell in existing Monte Carlo merger simulations, consider that in~\cite{Foucart:2022kon} the {\it minimum} packet energy is $10^{-12}M_\odot c^2$ and the grid resolution is $\sim 200\,{\rm m}$. Even without attempting to extract any angular resolution from the simulation, the distribution function of $\sim 20\,{\rm MeV}$ neutrinos in any cell would then be evaluated as either $f=0$ or $f\gtrsim1$. The lowest energy bin in that simulation has $\epsilon_1=4\,{\rm MeV}$, so that at best we know $f_\nu$ to $\sim 10^3$ in that energy bin. After merger, the energy of electron antineutrino packets rises by two order of magnitude, making the situation worse. Things are slightly better for simulations that do not attempt to evolve a remnant neutron star (e.g.~\cite{Miller:2019dpt,Kawaguchi:2024naa}), as discussed below, though still not satisfactory if the objective is to estimate $f_\nu$.

The core of the problem with current weighting schemes for neutrinos, at least as far as our wish to resolve $f_\nu$ is concerned, should now become clear. Most simulations emit packets that represent a fixed number of neutrinos or a fixed total energy, at least within a given region of spacetime. In the regions from which neutrinos mainly originate, most of the neutrinos are in high energy bins (tens of MeV). The lower energy neutrinos, which are most likely to escape and interact with matter outflows, are then woefully underresolved (again, only if computing $f_\nu$; the absolute error in neutrino energy / number does not suffer from that error). Even for simulations that may have enough packets in a grid cell to resolve $f_\nu$ (many do not), the choice of weights implies very large errors in $f_\nu$ for low-energy neutrinos, and a clear breakdown of the $f_\nu<1$ constraints. We will quantify these issues more carefully in Sec.~\ref{sec:accuracy}.

\subsection{Need for the distribution function in simulations}

Before looking at more specific accuracy requirements, it is worth briefly discussing what an improved knowledge of the distribution function would allow, and where not knowing $f_\nu$ may already be a problem for existing simulations. One aspect of course is the desired to satisfy the exclusion principle, so that neutrinos are properly treated as fermions. Having $f_\nu\gg1$ in many regions of phase space in current simulations is not reassuring, even if $f_\nu<1$ when averaging over regions of phase space in which enough packets are available. The calculation of neutrino-matter and neutrino-neutrino interactions is however a more significant motivation.

Current merger simulations mainly aim to include three broad types of reactions: neutrino creation/annihilation through interactions with particles in thermal equilibrium, e.g. $p+e^- \leftrightarrow n + \nu_e$, scattering of neutrinos on other particles, e.g. $n + \nu \rightarrow n + \nu$ and thermal production and annihilation of pairs of (anti)neutrinos, e.g. $e^+ + e^- \leftrightarrow \nu_\mu \bar\nu_\mu$. In terms of the evolution of $f_\nu$, we can write each as
\beq
\partial_t f_\nu = \Gamma^+ - \Gamma^-,
\eeq
for creation and annihilation rates $\Gamma^\pm$. As non-neutrino particles are assumed to be in statistical equilibrium in the fluid, the part of these rates that cannot be calculated from statistical mechanics and nuclear physics are terms proportional to $f_\nu$ when reactions destroy neutrinos, and $(1-f_\nu)$ when reactions create neutrinos (blocking factors).

For reactions creating/destroying a single neutrino, we thus have
\beq
\Gamma^+ = A (1-f_\nu);\,\, \Gamma^- = B f_\nu,
\eeq
with $A,B$ coefficients that do not depend on $f_\nu$. This is obviously mathematically equivalent to
\beq
\Gamma^+ = A;\,\, \Gamma^- = (A+B) f_\nu,
\eeq
a formulation that avoids the use of blocking factors in simulations. $A$ provides us with a probability to emit packets, and $(A+B)$ a probabily to absorb packets. Using the equilibrium condition $\Gamma^+=\Gamma^-$ when $f_\nu = f_{\rm eq}$, with $f_{\rm eq}$ the Fermi-Dirac distribution of neutrinos in equilibrium with the fluid, one can also rewrite $(A+B)$ as a {\it stimulated absorption} coefficient (see e.g.~\cite{Burrows2006b}) that can be calculated directly from the absorption probability and the known $f_{\rm eq}$.

For scattering events, we have
\beqn
\Gamma^+ &=& (1-f_\nu(\epsilon,\Omega)) \int d\epsilon' d\Omega' f_\nu (\epsilon',\Omega') A(\epsilon',\epsilon,\mu),\nonumber\\
\Gamma^- &=&  f_\nu(\epsilon,\Omega)  \int d\epsilon' d\Omega' (1-f_\nu (\epsilon',\Omega')) A(\epsilon,\epsilon',\mu),
\eeqn
with $(\epsilon,\Omega)$ the energy and direction of the neutrinos, $\mu=\cos\theta$, and $\theta$ the angle between the momenta of the two neutrinos. Again, $A(\epsilon,\epsilon',\mu)$ does not depend on $f_\nu$. We have the same coefficient $A$ in both rates because this process does not create or destroy neutrinos; neutrinos `destroyed' by $\Gamma^-$ are `created' by $\Gamma^+$ with a different momentum. An important special case is elastic scattering, when $\epsilon=\epsilon'$, i.e. $A(\epsilon,\epsilon',\mu) = A(\epsilon,\mu) \delta(\epsilon-\epsilon')$. In that case, we can use the mathematically equivalent rates
\beqn
\Gamma^+ &=& \int d\Omega' f_\nu (\epsilon,\Omega') A(\epsilon,\mu),\nonumber\\
\Gamma^- &=&  f_\nu(\epsilon,\Omega)  \int d\Omega'  A(\epsilon,\mu).
\eeqn
This can be treated in Monte Carlo as a rate of scattering for any packet ($\Gamma^-$) and a probability distribution for the change of direction of the packet after scattering, without ever requiring calculations of $f_\nu$. For inelastic scattering, on the other hand, no such cancellation occurs. Elastic scattering is a good approximation for scattering on particles with rest mass energy much larger than $\epsilon_\nu$, which includes nucleons and nuclei in merger simulations but not electrons. Accordingly, existing merger simulations generally ignore scattering on electrons.

Finally, for pair processes,
\beqn
\Gamma^+ &=& (1-f_\nu(\epsilon,\Omega)) \int d\epsilon' d\Omega' (1-\bar f_\nu (\epsilon',\Omega')) A(\epsilon',\epsilon,\mu), \nonumber\\
\Gamma^- &=& f_\nu(\epsilon,\Omega)  \int d\epsilon' d\Omega' \bar f_\nu (\epsilon',\Omega') B(\epsilon,\epsilon',\mu),
\eeqn
with $\bar f_\nu$ the distribution of antiparticles. Here the situation is even worse, as even the raw absorption rate $\Gamma^-$ requires knowledge of $\bar f_\nu$. This is probably where current simulations most suffer from not knowing $f_\nu$. In particular, simulations that do not include charged-current reactions for muon and tau neutrinos rely solely on pair production as a source of heavy-lepton neutrinos. A common approximation is to estimate blocking factors in $\Gamma^+$ using the equilibrium distribution function, then calculate $\Gamma^-$ in such a way that $\partial_t f=0$ when $f=f_{\rm eq}$. This works as an approximate way to get the correct distribution in regions where neutrinos are trapped, but is inaccurate otherwise.

Inelastic scattering is expected to be important to the thermalization of heavy-lepton neutrinos~\cite{Cheong:2024buu}, pair production to the luminosity and spectrum of heavy-lepton neutrinos, and pair annihilation to the deposition of energy in low-density regions~\cite{Fujibayashi:2017puw}. All of these processes would greatly benefit from improved knowledge of $f_\nu$.

Flavor conversions, as already mentioned, also have a complex dependence in $f_\nu$. Even simulations that attempt to account for flavor conversions through subgrid models rather than through direct evolution of the quantum kinetics equation could make use of a more detailed knowledge of the angular distribution of neutrinos.

Overall, we thus see that merger simulations have so far gotten away with what would be extreme shot noise in their ability to resolve $f_\nu$ mostly thanks to the use of a limited set of reactions that can be written as source terms that are either constant or linear in $f_\nu$ -- at which point the fermionic nature of neutrinos can be hidden in the calculation of the emission and absorption coefficients rather than requiring explicit on-the-fly calculations of $f_\nu$. The exception are thermal processes of the type $e^+e^-\leftrightarrow \nu_\mu \bar\nu_\mu$, which are included in simulations in order to get realistic equilibrium distribution of heavy lepton nucleons while giving up on the calculation of accurate rates in other regimes.

\section{Requirements for specified accuracy}
\label{sec:accuracy}

\subsection{Desired weight of Monte Carlo packets}

Let us now consider a region of phase space $D$ within which the true average distribution function of neutrinos is $f_{\rm true}$. The expectation value for the total number of packets within $D$ is
\beq
\langle \mN \rangle = \frac{E_{\rm max} f_{\rm true}}{e_p},
\eeq
with $e_p$ the energy of each Monte Carlo packet (assumed to be the same for all packets here, for simplicity). 
When assessing the accuracy with which $f_\nu$ is computed, there are two distinct types of errors than we should consider: the purely statistical error due to the variance in the value of $N(D)$ in Monte Carlo simulations, and the numerical errors due to the finite resolution of the simulation itself (i.e. grid spacing used for the evolution of the  fluid and spacetime metric, time stepping algorithm and time step, interpolation of background variables to the location of a packet). In this manusript, we are interested in the former -- and, even more restrictively, solely in the number of packets required to limit shot noise errors in the derived values of $f_\nu$. The latter however dominates the error budget in many regime. For example, a slight error in the temperature of the fluid can lead to variations in $f_\nu$ far larger than the statistical error in the Monte Carlo algorithm, while an error in the propagation of neutrinos can lead to systematic biases in the location of the packets far from the remnant. These errors cannot be improved upon simply by adding more packets, however; and estimating them requires performing detailed convergence tests on full merger simulations.

The determination of the variance in the number of packets $\sigma_\mN$ in Monte-Carlo simulations is a non-trivial problem. When $\langle \mN\rangle \gg1$, a reasonable approximation is $\sigma_\mN \sim \sqrt{\mN}$. We discuss the origin and limitations of that approximation in more details in the Appendix. Here, we simply note that while it is a convenient method to estimate the number of packets needed to evaluate $f_\nu$, it is not a good substitute to full convergence studies if trying to evaluate errors in actual observables. When $\langle \mN \rangle \lesssim 1$, on the other hand, $\sigma_\mN \lesssim 1$ is more reasonable.

In the first case, the variance $\sigma_f$ in $f$ is approximately
\beq
\sigma_f = \frac{\sqrt{\mN}e_p}{E_{\rm max}} = \sqrt{\frac{f_{\rm true} e_p}{E_{\rm max}}}.
\eeq
We can evaluate the desired energy of Monte Carlo packets $e_p$ required to reach a target error $\sigma_f$ in the distribution function as
\beq
e_p = \frac{E_{\rm max} \sigma_f^2}{f_{\rm true}} = E_{\rm true} \left(\frac{\sigma_f}{f_{\rm true}}\right)^2,
\eeq
with $E_{\rm max}$ calculated as in the previous section and $E_{\rm true}$ the energy of neutrinos within $D$. The corresponding desired number of packets in $D$ is simply
\beq
\mN_{\rm target} = \left(\frac{f_{\rm true}}{\sigma_f}\right)^2,
\label{eq:Nt1}
\eeq
which is independent of grid resolution. 

When $\langle \mN \rangle \lesssim 1$, we conservatively estimate $\sigma_\mN\sim 1$ and get
\beq
\sigma_f \sim \frac{e_p}{E_{\rm max}} \rightarrow \mN_{\rm target} = \left(\frac{f_{\rm true}}{\sigma_f}\right).
\eeq

It should be noted that when using a fixed $\sigma_f$, we assume that we only need $f$ for the calculation of blocking factors, and thus care about absolute errors in $f$. This is shown by the fact that the desired average number of packets in $D$ is actually less than $1$ if $f_{\rm true}<\sigma_f$. For applications in which we desire a fixed relative error in $f$ (e.g. instabilities in neutrino flavor oscillations), one could pick a constant $\sigma_f/f_{\rm true}$ instead of a constant $\sigma_f$. 

Combining these two estimates, we get
\beq
e_p =E_{\rm true} \min{\left[\left(\frac{\sigma_f}{f_{\rm true}}\right)^2,\frac{\sigma_f}{f_{\rm true}}\right]}.
\eeq
and
\beq
\mN_{\rm target} = \max{\left[\left(\frac{f_{\rm true}}{\sigma_f}\right)^2,\frac{f_{\rm true}}{\sigma_f}\right]}.
\label{eq:Nt2}
\eeq
For the simple scaling formulae below, we use Eq.~\ref{eq:Nt1}, but for more detailed calculations we use Eq.~\ref{eq:Nt2}. This distinction is not a particularly important concern for the estimates of the required number of packets presented here, as those are dominated by regions where $\mN_{\rm target}\gg 1$ and Eq.~\ref{eq:Nt1} approximately holds; but it may be an important consideration if using such a prescription in actual simulations, to avoid packets with $e_p \gtrsim E_{\rm max}$.

Scaling to a relatively modest requirement of $\sigma_f \sim 0.1$, we get
\beqn
e_p &=& \frac{2.35\times 10^{-15} M_\odot c^2}{f_{\rm true}} \times\nonumber \\
&&\,\, \frac{\Delta \Omega}{4\pi} \frac{\Delta V}{(100\,{\rm m})^3} \frac{\epsilon_1^4-\epsilon_0^4}{(20\,{\rm MeV})^4} \left(\frac{\sigma_f}{0.1}\right)^2.
\label{eq:Ep}
\eeqn
The same calculation for packets representing a constant number of neutrinos $n_p$ gives a desired number of neutrinos per packets of
\beq
n_p = \frac{1.76\times 10^{44}}{f_{\rm true}} \frac{\Delta \Omega}{4\pi} \frac{\Delta V}{(100\,{\rm m})^3} \frac{\epsilon_1^3-\epsilon_0^3}{(20\,{\rm MeV})^3}\left(\frac{\sigma_f}{0.1}\right)^2.
\label{eq:Np}
\eeq

From these results, it should be clear that if the objective is to capture $f_\nu$ with a desired absolute accuracy, the optimal weight of packets is neither constant number nor constant energy, but rather a constant $\sigma_f$ across all bins and grid cells, i.e.
\beq
n_p \propto \frac{\epsilon_1^3-\epsilon_0^3}{f_{\rm true}}.
\eeq
This requires many more packets to be produced for low-energy neutrinos than in existing simulations, and much less packets for high-energy neutrinos. Whether such a scheme is desirable in practice is unclear: it would result in Monte Carlo packets for high-energy neutrinos that each represent a larger energy / number of neutrinos, impacting shot noise when coupling the transport equations to the evolution of the fluid's temperature and composition. High energy neutrinos are mostly trapped in the dense regions of the star, while the low-energy neutrinos that would be better resolved by such a choice are more likely to escape and interact with low-density regions; it is thus unclear whether such a choice would actually be hurtful to the accuracy of merger simulations. Careful testing is required to evaluate the feasibility of such a weighting scheme. The scheme additionally requires an estimate of $f_{\rm true}$; $f_{\rm eq}$ may be a reasonable choice here.

\subsection{Desired number of packets for specific fluid properties}

To illustrate accuracy requirements for $f_\nu$, we now consider neutrinos within a fluid at a given $(\rho,T,Y_e)$, and assess the number of packets that would be required within a domain $D$ to reach $\sigma_f=0.1$. For concreteness, we use the same binning in energy as in our neutron star merger simulations~\cite{Foucart:2022kon}, i.e. 16 energy bins with the lowest bin covering the range $[0,4]\,{\rm MeV}$, and other bins logarithmically spaced up to $528\,{\rm MeV}$. We also use the SFHo equation of state~\cite{2013ApJ...774...17S} to calculate equilibrium neutrino distributions, and 3 species of neutrinos: $\nu_e$, $\bar\nu_e$, and a $\nu_x$ species encompassing all other neutrinos (for that group, the estimates of $e_p$ and $n_p$ should be multiplied by 4, to account for the 4 species of neutrinos included in $\nu_x$). We note that this is not an optimal binning choice for our purpose here, and is presented mainly as a concrete choice consistent with what many simulations using the NuLib library of neutrino interactions would make~\cite{OConnor2015}. We assume no attempt to capture the angular dependence of the distribution function ($\Delta \Omega = 4\pi$). For other conditions, one can scale the desired number of packets using the relationships from the previous section.

We note that for simulations using 5-6 species of neutrinos, assuming about a factor of 2 increase in the number of required packets is a reasonable first approximation if the reactions included in the simulations do not change. If charged-current reactions involving heavy-lepton neutrinos are considered, a more detailed calculation would depend on the muon fraction in the fluid. The cost of using additional species may then be slightly higher: \cite{Ng:2024zve} for example finds that muon antineutrinos then become the dominant species in the densest regions just after merger, instead of electron antineutrinos.

We consider two potential packet weights: either constant $n_p$, or Eq.~\ref{eq:Nt2} for $\mN_{\rm target}$ in all bins. When using constant $n_p$, the bin with the strictest requirement sets $n_p$ for all bins, but we allow different $n_p$ for different neutrino species and different fluid parameters. Finally, we assume that within each energy bin
\beq
E_{\rm true} = \frac{\eta}{\kappa_a+\kappa_{floor}} \frac{\Delta V}{c},
\eeq
with $\eta$ the emissivity, $\kappa_a$ the absorption opacity, and $\kappa_{\rm floor}=c^2/GM_\odot$. Neutrinos in equilibrium with the fluid have $E_{\rm true} = \eta/\kappa_a$, and $\kappa_{\rm floor}$ is used to approximately limit $E_{\rm true}$ in regions of phase space in which neutrinos are free-streaming. We use the same reaction rates as in~\cite{Foucart:2022kon}.

We note that this is in some sense a generous approximation: it assumes that we can {\it locally} chooses $e_p$ to fit the fluid properties of a grid cell. In practice, neutrinos emitted in one region of the simulation with a locally-chosen $e_p$ may propagate to another region with a different desired $e_p$. This issue may be alleviated with adaptive packet weighting, i.e. either splitting or down-sampling packets when their current $e_p$ differs too much from the desired $e_p$. Such a scheme is already used in~\cite{Kawaguchi:2024naa}.

\begin{figure*}
\includegraphics[width=0.3\textwidth]{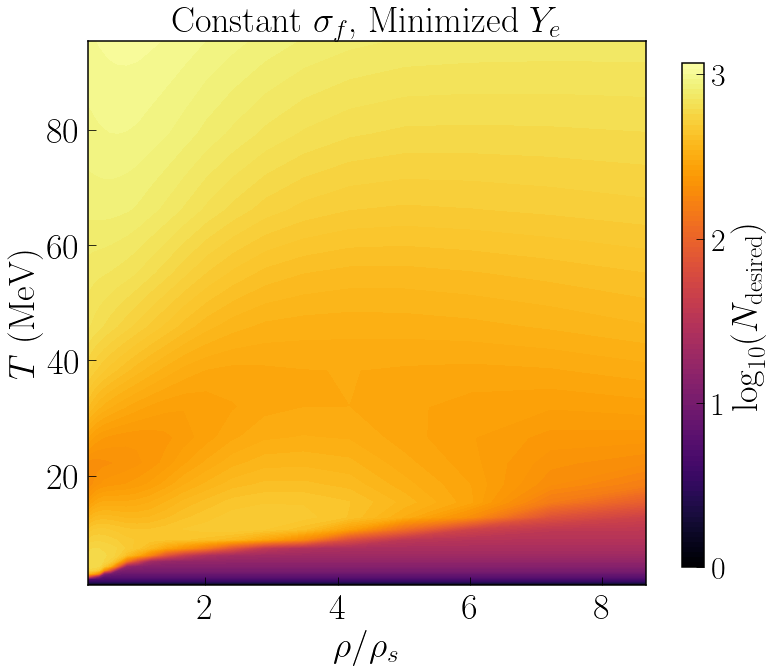}
\includegraphics[width=0.3\textwidth]{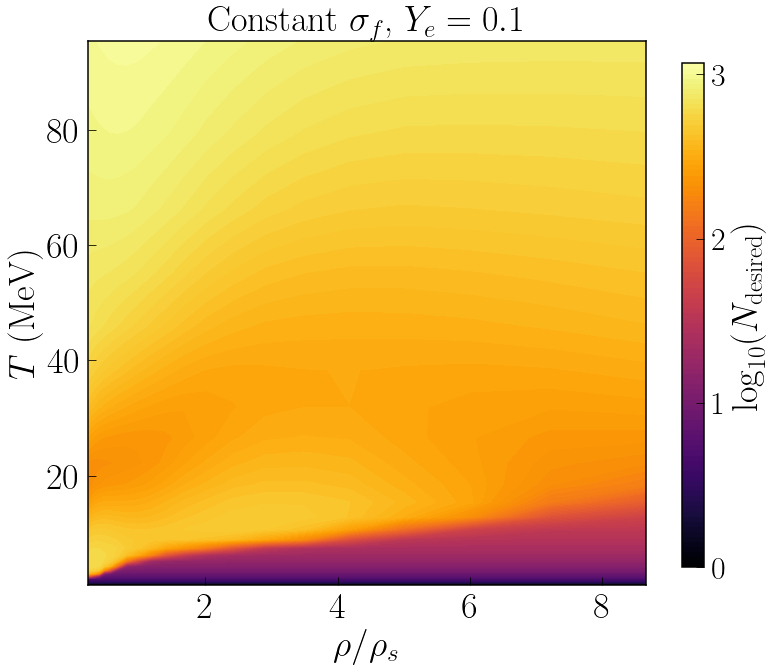}
\includegraphics[width=0.3\textwidth]{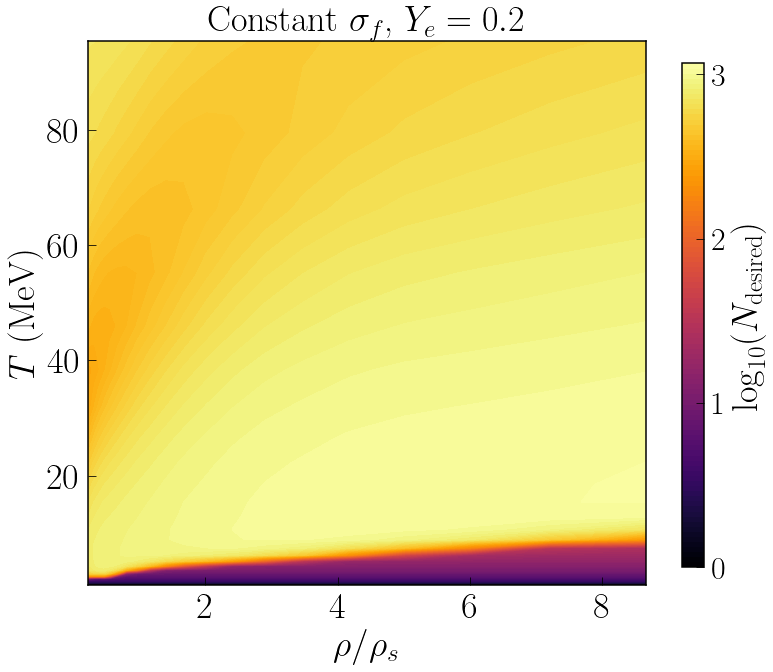}\\
\includegraphics[width=0.3\textwidth]{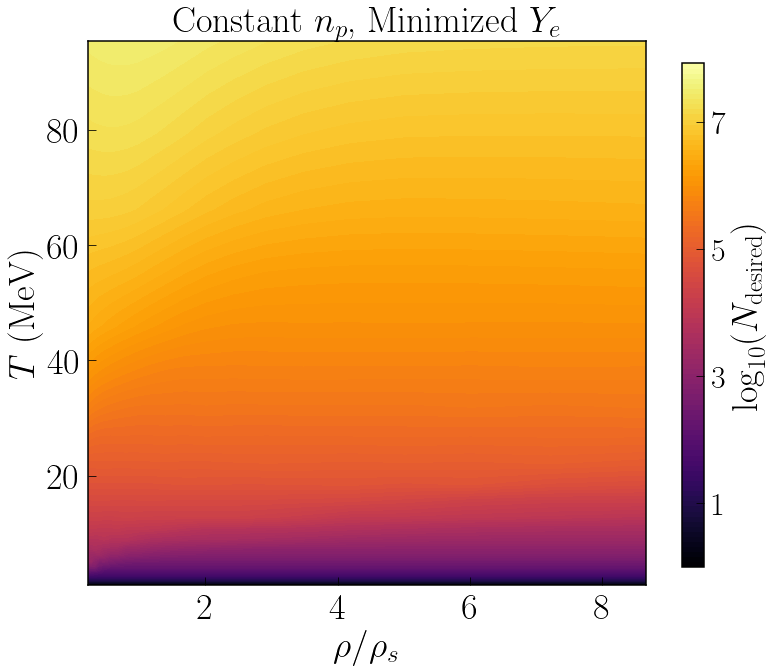}
\includegraphics[width=0.3\textwidth]{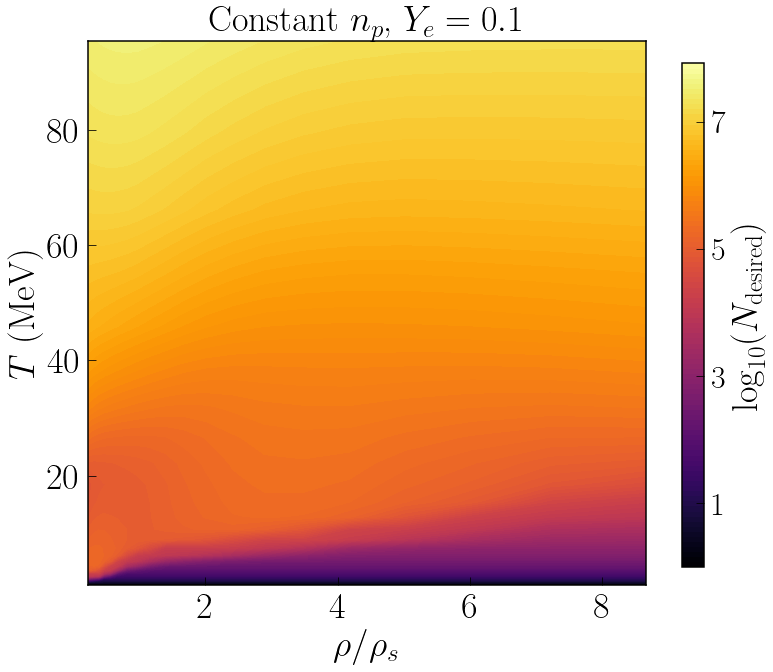}
\includegraphics[width=0.3\textwidth]{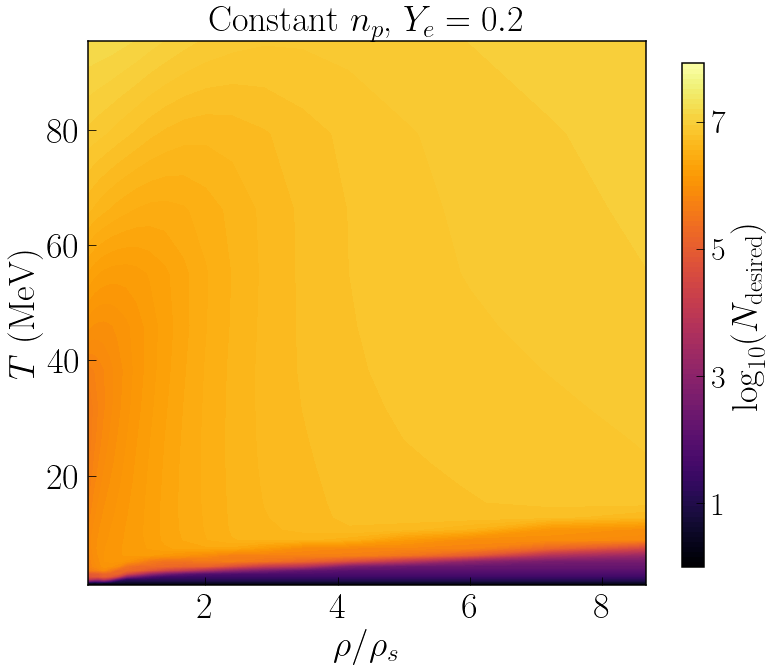}
 \caption{{\it Top}: Desired mean number of packets per cell assuming a desired $\sigma_f=0.1$ in each cell (or packet energy $e_p=\sigma_f E_{\rm max}$ when $f_\nu < \sigma_f$). The x-axis shows the baryon density normalized by the nuclear saturation density; the y-axis shows the temperature in MeV. {\it Bottom}: Same as Top, but using the same number of neutrinos per packet in all energy bins. Note the difference in scale when compared to the top plots. {\it Left}: Number of packets for the value of $Y_e$ that minimizes the packet count. {\it Center}: Number of packets for $Y_e=0.1$. {\it Right}: Number of packets for $Y_e=0.2$.}
\label{fig:high_rho}
\end{figure*}

\begin{figure*}
\includegraphics[width=0.3\textwidth]{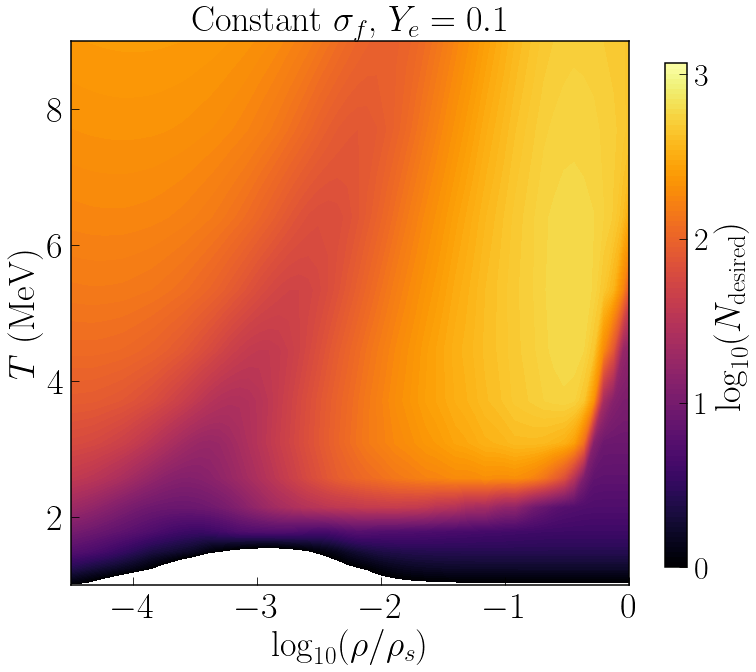}
\includegraphics[width=0.3\textwidth]{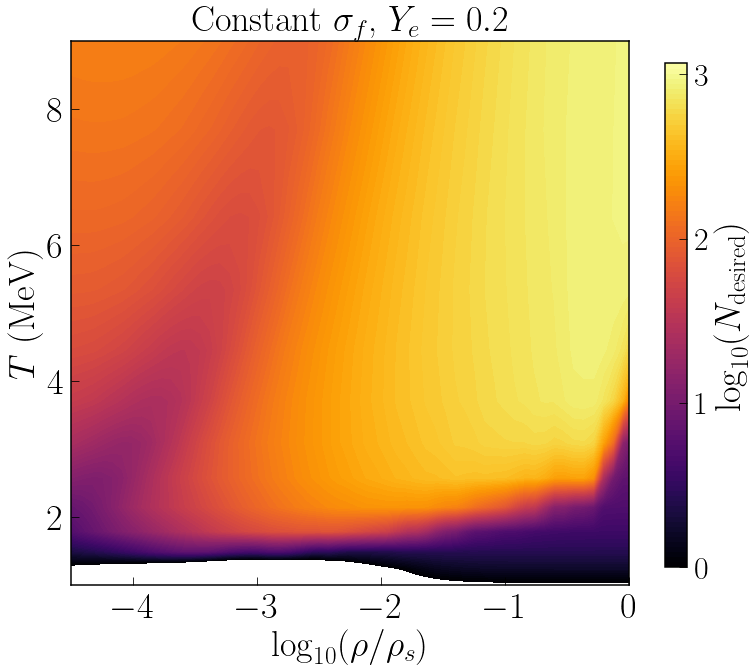}
\includegraphics[width=0.3\textwidth]{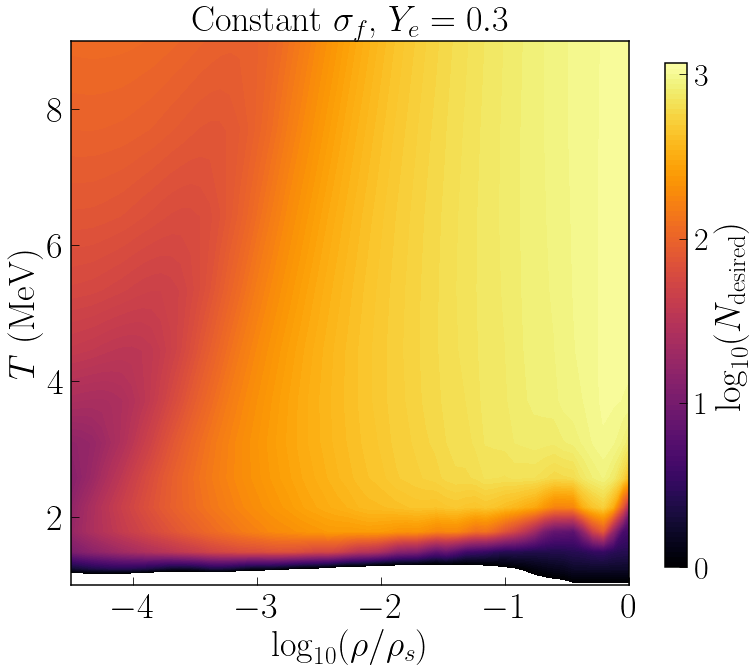}\\
\includegraphics[width=0.3\textwidth]{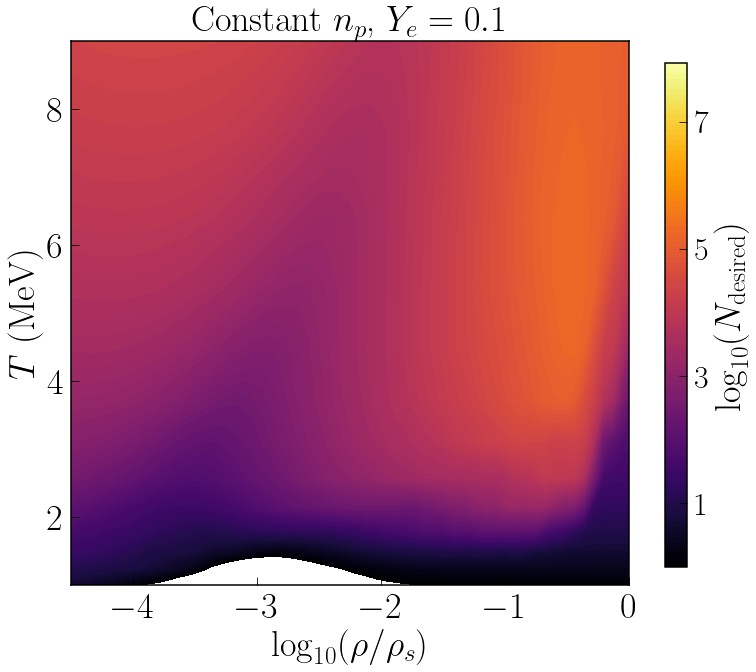}
\includegraphics[width=0.3\textwidth]{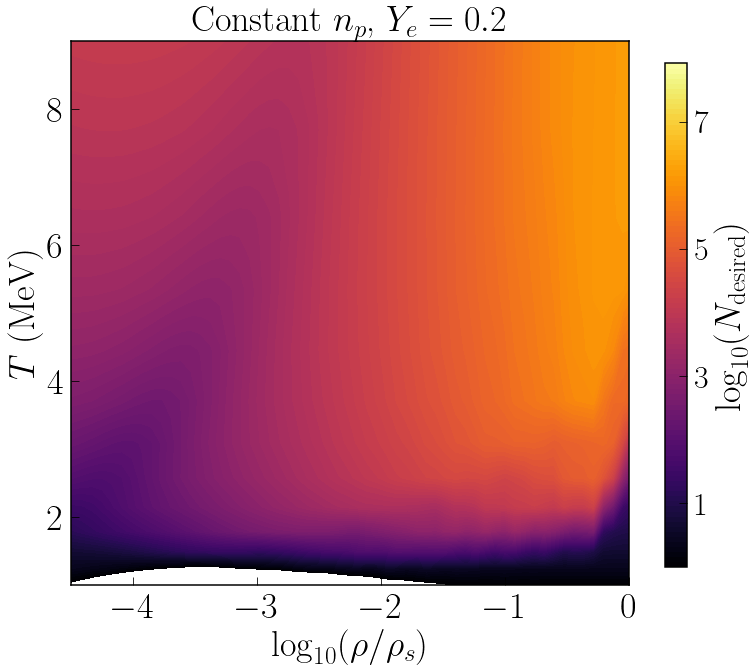}
\includegraphics[width=0.3\textwidth]{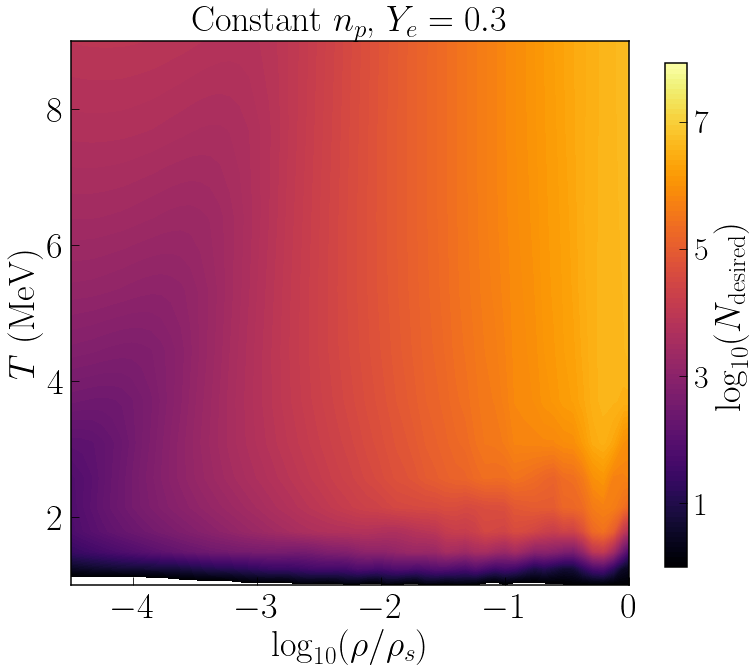}
 \caption{Same as Figs~\ref{fig:high_rho}, but for densities below nuclear saturation density and considering a fluid at a fixed $Y_e=0.1,0.2,0.3$. As before, note the use of different color scales for the top and bottom panels. {\it Top:} As in the top panel of Fig.~\ref{fig:high_rho}, we use $\sigma_f=0.1$ in all energy bins. {\it Bottom}: As in the bottom panel of Fig.~\ref{fig:high_rho}, we require the same number of neutrinos per packet in each energy bin. {\it Left to Right}: Values for $Y_e=0.1,0.2,0.3$. These conditions are more typical of the atmosphere of post-merger neutron stars (close to nuclear saturation density) and surrounding accretion disk ($\log_{10}(\rho/\rho_s)\lesssim -3$). Results for higher values of $Y_e$ are likely mostly relevant at lower densities in neutron star mergers.}
\label{fig:low_rho}
\end{figure*}

We consider two different sets of thermodynamical conditions. First we look at densities and temperatures typical of the inner regions of post-merger neutron star remnants (Fig~\ref{fig:high_rho}), then at conditions more relevant to the surface of neutron star remnants and to post-merger accretion disks (Fig.~\ref{fig:low_rho}). At high density, we show both the value of $Y_e$ minimizing the desired number of packets, which is very close to the pre-merger value of $Y_e$ in neutron star mergers ($Y_e<0.1$), as well as larger values of $Y_e=0.1,0.2$. At low densities, we show $Y_e=(0.1,0.2,0.3)$.

If all energy bins have $f_{\rm true}=1$, we would expect to need $100$ packets for each of the 16 energy bins and 3 neutrino species, i.e. $4800$ packets. In practice, not all bins have $f_{\rm true}=1$, even at high density and temperature, but in hot regions inside neutron stars we can reasonably expect that $O(1000)$ packets is close to optimal for our choice of $\sigma_f=0.1$, 3 neutrino species, and $16$ energy bins. The top panel of Fig.~\ref{fig:high_rho}, which simply uses Eq.~\ref{eq:Nt2} to independently set the number of packets in each energy bin, confirms that this is indeed the case for $T\gtrsim 50\,{\rm MeV}$. In colder regions, we need at least 100 packets per cell. Moving away from the `best case' electron fraction does not qualitatively change the results much when using Eq.~\ref{eq:Nt2}. The results at $Y_e=0.1$ are very similar to the best-case scenario. Cold regions with $Y_e=0.2$ would require a significantly larger number of packets, but we do not expect such regions to exist in neutron star mergers: only hot regions undergo such large changes of composition.

The bottom panel of Fig.~\ref{fig:high_rho} shows the same thermodynamical conditions, but setting $n_p$ in all energy bins to a shared value low enough to get $\sigma_f=0.1$ in all bins. This is closer to the method used currently in merger simulations\footnote{Existing merger simulations use a fixed $e_p$ in each cell, which leads to desired numbers of particle about an order of magnitude higher; using a fixed $n_p$ would be a fairly simple modifications, however.}, but leads to a much higher required number of packets. This stricter requirement comes from the fact that $n_p$ is tyically set to a small number needed to get $\sigma_f \sim 0.1$ in the lowest energy bin, leading to a very large number of packets in the higher energy bins. Noting the different color scale used in the bottom panel of Fig.~\ref{fig:high_rho}, we see that we now require $10^{5-7}$ packets per cells in large regions of phase space. This would be completely impractical in numerical simulations: existing merger simulations using Monte Carlo methods use $\sim 10^8$ packets over the entire computational domain, and have $\sim 10^6$ cells covering the inside of the post-merger remnant.

Overall, we thus see that while resolving the distribution function with $\sigma_f\sim 0.1$ may not be entirely out of reach for merger simulations in theory, it would at the very least require careful revisions of the weighting scheme, and likely the use of additional approximations in the calculation of the distribution function (e.g. evaluating $f_\nu$ over larger regions of phase space than the discretization of the simulation, see Discussion section). The unknown impact of a rather extreme weighting scheme for neutrino packets on the coupling between the fluid and the neutrinos may complicate its use in practice.

At lower densities, and especially for values $\rho<10^{-3}\rho_s$ typical of post-merger accretion disks (with $\rho_s$ the nuclear saturation density), the situation is generally better -- though direct use of the distribution function in simulations would still require significant care. We illustrate this in Fig.~\ref{fig:low_rho} using the same methods to calculate the number of packets as in Fig.~\ref{fig:high_rho}, but at lower densities and temperatures $T<10\,{\rm MeV}$. As matter at lower density will experience more significant changes to its composition than the core of neutron stars, we now span a wider range of $Y_e=0.1,0.2,0.3$. We see that, as before, $\sim 1000$ packets are required close to nuclear saturation density when using the `optimal' choice from Eq.~\ref{eq:Nt2}, but $>10^5$ packets when fixing $n_p$ instead. At lower densities more typical of post-merger accretion disks, this is reduced to $\sim 100$ packets only when using Eq.~\ref{eq:Nt2}, and $\sim 10^4$ packets when using a constant $n_p$. Overall, in a post-merger accretion disk, it may thus be possible to coarsely capture the energy dependence of $f_\nu$ with more moderate changes to the weighting of the neutrino packets than in merger simulations: the axisymmetric simulations of~\cite{Kawaguchi:2024naa} have an effective number of packets available for calculation of $f_\nu$ of $\sim 1000$.

\section{Discussion}

From the above discussion, we see that to realize the promises of Monte Carlo to capture the energy and angular dependence of neutrinos in neutron star merger simulations, we would at the very least require significant changes to the way we weight neutrino packets, and likely some smoothing in the calculations of $f_\nu$ on scales larger than the grid scale (in space) and energy binning scale. We provide here simple guidelines for the conditions that need to be met in order to meet a specified accuracy in $f_\nu$. Eqs.~\ref{eq:Ep}-\ref{eq:Np}, for example, can be used either to set requirements on the desired weight of Monte Carlo packets for a given resolution, or to assess the size of the domain of phase space that needs to be considered in order to get reasonable estimates of $f_\nu$ for a chosen value of the weights.

Our results clearly show that using either a fixed number of neutrinos or a fixed total energy for each packets within a region of spacetime inevitably makes it very costly to obtain even coarse estimates of $f_\nu$ for low energy neutrinos. This is true even if the weights are allowed to vary with the time and position of the packets. If estimating $f_\nu$ is an important target of a simulation, low-energy neutrinos have to be given more weight than high-energy neutrinos (or have to be more coarsely resolved) -- an objective that may run counter to our desire to limit shot noise in the coupling between the evolution of the neutrinos and that of the fluid's temperature and composition. Even with weights chosen to minimize errors in $f_\nu$, errors $\sigma_f \sim 0.1$ require $O(100)$ packets within any energy or angular bin where $f_\nu\sim 1$, a costly requirements considering that existing 3D merger simulations use at most $O(100)$ packets per grid cell.

A few different avenues may be consider in order to move forward with estimates of $f_\nu$, and the inclusion of reactions such as pair processes and inelastic scattering. The simplest is of course to continue to use $f_{\rm eq}$ whenever the distribution function is required (or maybe $f\rightarrow 0$ when neutrinos are free-streaming), but this will significantly limit the physical realism of merger simulations. A more satisfactory answer is likely to involve a choice of packet weights more adapted to calculations of $f_\nu$ combined with a smoothing of $f_\nu$. For existing Monte Carlo simulations with grid spacing of $\sim 200\,{\rm m}$, averaging $f_\nu$ over multiple grid cells is not very appealing, as a typical lengthscale for changes in the properties of the fluid is $\sim 1\,{\rm km}$. Future simulations combining neutrinos and magnetic fields will necessarily require finer grid resolution. By that point, averaging over all immediately neighboring cells becomes easier to justify, and would provide $\sim 27$ times more packets for the calculation of $f_\nu$ without modifying the relative cost of radiation transport in the simulations. With ideal weighting of the neutrino packets, getting $\sigma_f \lesssim 0.1$ may then become realistic for Monte Carlo transport, with the cost of radiation transport remaining comparable to the cost of the fluid evolution (as is currently the case in our merger simulations).

Averaging in energy space may also be a useful way forward. In particular, a weighting scheme optimized for calculations of $f_\nu$ would use a significant fraction of our computational resource to evolve very low energy neutrinos ($<5\,{\rm Mev}$) that are not as important to the evolution of the merger remnant than $(10-20)\,{\rm MeV}$ neutrinos. It may be valuable to merge low-energy bins for the purpose of calculating $f_\nu$ at least.
This may prove particularly useful in dense, hot regions, where our scheme enforcing constant $\sigma_f$ may waste significant resources resolving a large number of low-energy neutrino packets that are in equilibrium with the fluid.

It is also worth nothing that in regions of phase space where neutrinos quickly equilibrate with the fluid, assuming $f_\nu = f_{\rm eq}$ is a more accurate approximation than calculating $f_\nu$ from available neutrino packets for basically any number of packets that would be realistic today. There is thus little incentive to try to capture $f_\nu$ from the Monte Carlo packets in those regions. As equilibrium regions are easily identifiable, we may preemptively ignore any requirement coming from the need to estimate $f_\nu$ when in such a region. Methods combining moment methods in regions where neutrinos are trapped with Monte Carlo methods in low-density regions may particularly benefit from this, as the moment schemes will be primarily relied upon in hot, dense regions. The weighting of the Monte Carlo packets can then focus on capturing neutrino effects in regions where neutrinos are partially decoupled from the fluid, which are generally less costly for Monte Carlo simulations.

The road towards calculations of $f_\nu$ in Monte Carlo simulations thus appears difficult but not hopeless, even in the short term. And it may be worth considering if we find simulations that implicitly have quantum states occupied by $O(10^5)$ fermions to be mildly worrisome.

\begin{acknowledgments}
F.F. thanks the participants to the {\it Solving the Boltzmann Equation for Neutrino Transport in Relativistic Astrophysics} workshop at ICERM for discussions that partially motivated this work.
F.F. gratefully acknowledges support from the Department of Energy, Office of Science, Office of Nuclear Physics, under contract number DE-SC0020435 and from NASA through grant 80NSSC22K0719. 
\end{acknowledgments}

\appendix

\section{Additional comments on errors in Monte Carlo transport}

In the estimates of statistical errors presented in this work, we assumed that if the expectation value of the number of packets in a region of phase space is $\langle \mN \rangle$, then the variance is $\sigma_\mN \sim \sqrt{\langle \mN \rangle}$ for $\langle \mN \rangle \gg 1$ and $\sigma_\mN \lesssim 1$ for $\langle \mN \rangle \lesssim 1$. This is however an approximation. We discuss its origin and limitations in more details here.

First, as mentioned in the main text, it is crucial to note that we are only trying to account for the statistical error in the Monte Carlo simulation, i.e. assuming no error in the evolution of the fluid and spacetime variables, in the interpolation of these variables to the location of the Monte Carlo packets, or in the propagation of the packets themselves. Or, maybe more accurately, we want to know how many packets are needed to avoid ``shot noise'' errors in the calculation of the distribution function $f_\nu$. Propagation errors are typically small over the short times that packets spend on the computational grid. They may however lead to systematic biases in the location and/or four-momentum of the neutrinos. For sharply peaked distribution (in any dimension), this can easily lead to large errors in local estimates of $f_\nu$. A simple example would be a narrow beam evolved in curved spacetime. A small error in the trajectory of the beam would quickly lead to O(1) local errors in $f_\nu$. This is true even if the simulation is, by most standards, very accurate. Errors in the evolution of the temperature and composition of the fluid can also lead to O(1) errors in $f_\nu$, given the sensitivity of the neutrino distribution function to the properties of the fluid. These are clearly important issues to keep in mind if one wants to estimate the error in $N(D)$ within any region $D$ of a simulation and extract observables from radiation transport simulations. Simulation results need to converge both with the number of packets used in a simulation, and with the spacetime discretization used to evolve the fluid and metric. The question that we are posing here is however different. We want to know how many packets are needed within a numerical simulation so that statistical errors in the reconstruction of $f_\nu$ do not limit our ability to use on-the-fly calculations of $f_\nu$ -- even if that value of $f_\nu$ is biased by other sources of errors. For example, in the above example of a slightly erroneous temperature or composition, one may still self-consistently calculate neutrino-matter interactions using $f_\nu$ as measured in the simulations, if we have a sufficient number of packets within the region $D$ over which we average $f_\nu$. This will not add significantly to the error in the simulation. On the other hand, if a single packet in $D$ leads to an estimate $f_\nu \sim 10^{3-5}$, as in some regions of existing merger simulations, using the measured $f_\nu$ to calculate interaction rates will create huge errors even if the expectation value of $f_\nu$ is unbiased.

Putting aside non-statistical errors, the choice $\sigma_\mN\sim \sqrt{\mN}$ used for the purely statistical errors when $\mN\gg 1$ is itself an approximation. It is most rigorously defined if the evolution of different packets can be considered as independent, and statistical noise in the evolution of the fluid is neglected. For any two small regions of phase space $D_i$ and $D_f$, one can then define an ensemble of potential packets that may be created in $D_i$\footnote{The exact definition will depend on how emission is sampled in practice, which varies in different implementations of Monte Carlo transport} and a probability $p$ that such a packet then passes through $D_f$. This is true regardless of how many interactions that packet undergoes between its emission and its passage through $D_f$. If, over the entire phase space of the simulation, there are a total of $\mN_{\rm tot}$ potential packets with probability $p$ to pass through $D_f$, then the expectation value for the number of packets that actually passes through $D_f$ is $\langle \mN \rangle = \mN_{\rm tot} p$, and the variance is $\sqrt{\mN_{\rm tot}p(1-p)}=\sqrt{\langle \mN \rangle(1-p)}$ for $\langle \mN \rangle\gg 1$ (for independent draws from a binomial distribution). Combining the variance of packets with different $p$ (assumed, again, independent), and considering that $(1-p) \sim 1$, one gets $\sigma_\mN\sim \sqrt{\mN}$.\footnote{Few packets will have $p$ close enough to 1 for this to be a bad approximation; though the existence of such packets only reduces $\sigma_\mN$.} Whether this arguments guarantees that $\sigma_\mN \sim \sqrt{\mN}$ is a good approximation or not then depends on how much correlations between the evolution of different packets impact that result. We can at least say that for an optically thin system with a relatively long timescale for cooling and composition evolution, e.g. an accretion disk, the approximation $\sigma_\mN=\sqrt{\mN}$ is on fairly solid ground.

In current simulations, correlations will be due to the fact that the evolution of a packet has an effect on the background fluid, which itself clearly impacts the emission and interaction probabilities of other packets. In that respect, $\sigma_\mN \sim \sqrt{\mN}$ is at least a good estimate for the variance in the number of packets in a region of phase space for a specific realization of the fluid background. This might seem to be a pretty significant caveat, but for the fact that existing Monte Carlo simulations show shot noise in the fluid variables largely limited to low-density regions in which few neutrinos are interacting. The main reason for this is that for systems like neutron star merger remnants where the energy density of neutrinos is much smaller than the energy density of the fluid, shot noise in the fluid evolution is due to interactions of the fluid with many Monte Carlo packets over long timescales. Shot noise in the fluid variables will then be limited by the fact that we are averaging neutrino-matter interactions over interactions with a large number of packets. It should be acknowledged however that we cannot guarantee that this is the case everywhere in the simulation.

More generally, our assumptions will certainly break down over timescales long enough for statistical noise to cause a drift in the evolution of the fluid variables from the true solution. Errors in the variation of the background fluid under the influence of neutrino-matter interactions should scale with the square root of the total number of packets {\it for a given choice of numerical methods, including packet weighting}. Like discretization error in the evolution of the fluid, however, these errors have no reason to lead to $\sigma_\mN \sim \sqrt{\mN}$ {\it locally}. These errors are also likely to be worse for weighting schemes that try to capture $f_\nu$ locally than for constant-$e_p$ or constant-$n_p$ weighting, due to the larger shot noise associated with the evolution of a small number of heavily weighted packets representing high-energy neutrinos in the former schemes. To estimate these errors, one needs to perform the same simulation with different number of Monte Carlo packets (see e.g.~\cite{Foucart:2020qjb}); they cannot be estimated directly from local considerations based on the number of packets. In some way, there are two distinct problems here: the determination of the number of packets needed to estimate $f_\nu$ locally for a given fluid evolution, and the global accuracy of the simulation over long timescales under the influence of neutrino-matter interactions. With the purely local prescriptions used here, we can only try to address the former. 

For direct interactions using a measured $f_\nu$, which are not currently included in simulations, there are reasonable heuristic arguments for $\sigma_\mN=\sqrt{\mN}$ to remain valid under the same caveats as in the previous paragraphs. If a region of phase space can be populated from $\mN_{\rm in}$ packets each having a probabilty $p$ to populate that region through some reaction, the variance in the number of packets created due to that process is $\sqrt{p\mN_{\rm in}(1-p)}$, while the variance due to the preexisting uncertainty $\sqrt{\mN_{\rm in}}$ in $\mN_{\rm in}$ is (approximately) $p\sqrt{\mN_{\rm in}}$. The latter dominates the error budget unless $(1-p) \ll 1 $. Similar arguments can be made for other processes in which a large number of packets have a probability to create a smaller number of packets in a different region of phase space. These arguments break when a single packet can be responsible for the creation of a larger number of packets in a different region of phase space (processes with an effective $p>1$, e.g. cascades). We do not have any physical process that would behave in this way for neutrinos in mergers. We do however have a numerical process that may: splitting Monte Carlo packets as they propagate to adapt to local requirements on the desired number of neutrinos $n_p$ represented by that packet. For the choices of $n_p$ described here, this would happen when packets move from coarse grid cells to fine grid cells (i.e. when $\Delta V$ decreases). The packets created as a result of that splitting process are clearly very strongly correlated, at least until they experience further interactions. Immediately after splitting, having more packets does not help reduce the shot noise in measurements of $f_\nu$, and this should be considered when estimating $\sigma_f$. A more detailed analysis of individual reactions would be sensitive to how creation and annihilation rates are practically implemented in simulations, and is certainly something that one should consider when such methods are developed.

Overall, we thus see that the $\sigma_\mN\sim \sqrt{\mN}$ prescription should be taken with a fair amount of caution. It is only rigorously defined under very narrow constraints that do not apply everywhere in neutron star merger simulations, and only accounts for a very specific source of error in the simulations. It is about as good of an estimate as one can do without full convergence tests in simulations, and is thus what we use to guide our generic discussion of the needs of merger simulations. It is however definitely not an adequate replacement for convergence tests if one aims to obtain reliable error estimates.

\bibliography{References/References.bib}

\end{document}